\documentclass[twocolumn,superscriptaddress,longbibliography,aps,
pra,preprintnumbers]{revtex4-1}

\usepackage{graphicx}
\usepackage{bm}
\usepackage{color}
\usepackage{epstopdf}
\usepackage{amsmath}
\usepackage{amssymb}
\usepackage{epstopdf}

\usepackage[urlcolor=blue,colorlinks=true,citecolor=blue,linkcolor=blue,pdfstartview={FitH},bookmarks=false]{hyperref}

\newcommand{\Tr}{\text{Tr}}

\graphicspath{{fig/}{./fig/}{.}}

\begin{document}

\title{Reentrant Fulde--Ferrell--Larkin--Ovchinnikov superfluidity in the honeycomb lattice}

\author{Agnieszka Cichy}
\email[e-mail: ]{agnieszkakujawa2311@gmail.com}
\affiliation{Solid State Theory Division, Faculty of Physics, Adam Mickiewicz University, \\Umultowska
85, 61-614 Pozna\'n, Poland}
\affiliation{Institut f\"{u}r Physik, Johannes Gutenberg-Universit\"{a}t Mainz, \\
Staudingerweg 9, D-55099 Mainz, Germany}

\author{Andrzej Ptok}
\email[e-mail: ]{aptok@mmj.pl}
\affiliation{Institute of Nuclear Physics, Polish Academy of Sciences, \\ ul. E. Radzikowskiego 152, PL-31342 Krak\'{o}w, Poland}

\date{\today}

\begin{abstract}
We study superconducting properties of population-imbalanced ultracold Fermi mixtures in the honeycomb lattice that can be effectively described by the spin-imbalanced attractive Hubbard model in the presence of a Zeeman magnetic field.
We use the mean-field theory approach to obtain ground state phase diagrams including the unconventional Fulde--Ferrell--Larkin--Ovchinnikov (FFLO) phase, which is characterized by atypical behavior of the Cooper pairs total momentum.
We show that the momentum changes its value as well as direction with change of the system parameters.
We discuss the influence of van Hove singularities on the possibility of the reentrant FFLO phase occurrence, without a BCS precursor.
\end{abstract}

\maketitle

\section{Introduction}

The discovery of graphene~\cite{novoselov.geim.04} triggered enormous theoretical and experimental activity~\cite{castroneto.guinea.09,kotov.uchoa.12}. Henceforth, it has attracted much attention due to theoretical interests in fundamental physics, as well as its potential practical applications~\cite{park.schendel.14,zhou.zheng.15,lee.choi.16,liu.debnath.16,cervetti.rettori.16}. 
The attempt to understand graphene physics is not without difficulties, related e.g.\ to electron-phonon interactions and the presence of a charge inhomogeneity~\cite{loon.10}. 
However, recent advances in experiments offer the possibility to simulate similar condensed matter phenomena by loading ultracold bosonic or fermionic atoms into optical lattices~\cite{zhao.paramekanti.06,zhu.wang.07,soltanpanahi.struck.11,tarruell.greif.12,gomes.mar.12,uehlinger.jotzu.13,polini.guinea.13,messer.desbuquois.15,vasic.petrescu.15}. 
The engineering of the honeycomb lattice in ultracold gases setups as well as the creation of artificial graphene-like band structures bring the possibility of exploration of regimes which are still inaccessible in solid state materials. 
Recently, condensed matter systems based on fermions with linear dispersion (e.g.\ the honeycomb lattice) have generated a surge of intensive studies~\cite{blackchaffer.doniach.07,lee.bouadim.09,chen.liu.12,
li.guo.14,blackschaffer.wu.14,beugeling.kalesaki.15,sadeddine.enriquez.17,grass.chhajlany.17}. 
These models have substantial differences from models with an extended Fermi surface, such as those on the square lattice. 
However, it has not been understood yet which unconventional phases can be stable in such systems, especially in those where effective attraction is dominant.

In this work, we analyze the stability of one of the most interesting phases occuring in this type of systems, the Fulde--Ferrell--Larkin--Ovchinnikov (FFLO) state (formation of Cooper pairs across the spin-split Fermi surface with non-zero total momentum)~\cite{FF,LO}.
We consider the attractive Hubbard model in presence of a Zeeman magnetic field. 
It is worth to mention that at half-filling, in the absence of a Zeeman magnetic field, a quantum phase transition from the BCS state to the normal phase takes place.  It results in the occurrence of a critical attraction below which the BCS state is unstable.
Our main finding is not only establishing that the FFLO phase is stable for a wide range of parameters, but also that reentrant FFLO superconductivity can occur.
Moreover, at half-filling and in the spin imbalanced system (equivalent to a non-zero Zeeman magnetic field), the presence of van Hove singularities (VHS) in the density of states results in a stable FFLO phase for arbitrarily weak attractive interactions.
This discovery is essential from the viewpoint of realizing the FFLO state in ultracold gases experiments~\cite{liao.rittner.10} in artificial hexagonal lattices.
The field of such experiments has matured over the past decade~\cite{bloch.dalibard.08,soltanpanahi.struck.11,singha.gibertini.11,tarruell.greif.12,gomes.mar.12,polini.guinea.13,uehlinger.jotzu.13,mayaffre.kramer.14,revelle.fry.16,kinnunen.baarsma.18}.
In particular, investigations of quantum Fermi gases with spin or mass imbalance have become very popular~\cite{zwierlein.schirotzek.06,
partridge.li.06,zwierlein.ketterle.06,sheehy.radzihovsky.06,
radzihovsky.sheehy.07,conduit.conlon.08,conduit.green.09,radzihovsky.sheehy.10}. 
The possibility to control population imbalance by preparing mixtures with arbitrary population ratios motivates attempts to understand the influence of a Zeeman magnetic field on superfluidity.

The FFLO phase can be stable at low temperature and relatively large magnetic field (above the critical magnetic field of the Clogston-Chandrasekhar or Pauli limit~\cite{chandrasekhar.62, clogston.62}). 
There are experimental and theoretical premises that the FFLO state can be found in {\it quasi}-2D organic~\cite{piazza.zwerger.16,lortz.wang.07,bergk.demuer.11,beyer.wosnitza.13}, heavy-fermion~\cite{bianchi.movshovich.02,radovan.fortune.03,bianchi.movshovich.03,watanabe.kasahara.04,matsuda.shimahara.07,ptok.kapcia.17} or iron-based~\cite{ptok.kapcia.17,cho.kim.11,zocco.grube.13,ptok.15,ptok.crivelli.13,cho.yang.17,kasahara.watashige.14} superconductors. 
In this type of materials, a first-order phase transition from the superconducting to the normal state has been reported~\cite{lortz.wang.07,bergk.demuer.11,beyer.wosnitza.13,bianchi.movshovich.02,bianchi.movshovich.03,matsuda.shimahara.07,cho.kim.11,zocco.grube.13,ptok.15,watanabe.kasahara.04}. 
However, the observation of this type of superconductivity is very difficult because of the strong destructive influence of the orbital (paramagnetic) effect in solid state systems~\cite{gruenberg.gunther.66,matsuda.shimahara.07}.

Bringing together the two important threads of research, one related to graphene and the honeycomb lattice and the second to population imbalance in ultracold atomic gases, can lead to new and interesting physics. In particular, it gives the possibility to investigate some exotic superconducting phases which could potentially be found experimentally. So far, such phases have eluded experimental realization and one of the reasons for it is the non-zero critical value of attraction for the existence of the standard superconducting phase in the honeycomb lattice at half-filing and without magnetic field~\cite{zhao.paramekanti.06,su.tan.09}.
We show that reentrant FFLO superconductivity can be realized even below this critical value (even for arbitrarily small attractions) for some range of magnetic fields.
This greatly facilities the experimental realization and detection of the FFLO phase in ultracold fermionic gases in the lattice and makes searches for such a phase realistic. As such, it is the main finding of our work.

The paper is organized as follows. Section~\ref{sec.model}  gives a discussion of the spin-polarized Hubbard model as well as the method. Section~\ref{sec.numres} presents numerical results concerning among others the phase diagram, density of states analysis and the dependence of the Cooper pairs properties. We conclude in Section~\ref{sec.summary}.

\section{Model and technique}
\label{sec.model}

\begin{figure}[!t]
\centering
\includegraphics[width=\linewidth]{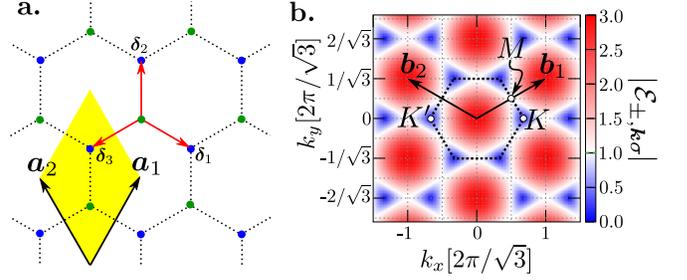}
\caption{
\label{fig.latt}
Panel a: honeycomb lattice discussed in this paper.
The unit cell defined by vectors $\bm{a}_1$ and $\bm{a}_2$ containing two atoms (blue and green points) belonging to sublattices A and B.
The three nearest-neighbor vectors are given by ${\bm \delta}_{1} = ( \sqrt{3}/2 , -1/2 )$, ${\bm \delta}_{2} = ( 0 , 1 )$ and ${\bm \delta}_{3} = ( -\sqrt{3}/2 , -1/2 )$.
The elementary primitive unit cell (yellow rhombus) is given by lattice vectors: ${\bm a}_{1} = {\bm \delta}_{2} - {\bm \delta}_{3} = a ( 1/2 , \sqrt{3}/2 )$ and ${\bm a}_{2} = {\bm \delta}_{2} - {\bm \delta}_{1} = a ( - 1/2 , \sqrt{3}/2 )$, where $a = \sqrt{3}$.
Panel b: dispersion relation for $\mu/t = 0$ and $h/t = 0$. 
Reciprocal lattice vectors are given by ${\bm b}_{1} = 2 \pi / a ( 1 , 1/\sqrt{3} )$ and ${\bm b}_{2} = 2 \pi / a ( -1 , 1/\sqrt{3} )$, while the first Brillouin zone is shown by a black doted hexagon.
High symmetry points are given by $M = {\bm b}_{1} / 2 = 2 \pi / a ( 1/2 , 1/2\sqrt{3} )$ and $K = - K' = ( {\bm b}_{1} - {\bm b}_{2} ) / 3 = 2 \pi / a ( 2/3 , 0 )$.
}
\end{figure}

The system can be described by the Hamiltonian in real space as $H = H_{K} + H_{I}$, where:
\begin{eqnarray}
\label{eq.hkin} H_{K} = \sum_{ i,j,s,s' \sigma} \left( - t_{ij}^{ss'} - \left( \mu + \sigma h \right) \delta_{ij} \delta_{ss'} \right)c_{i s \sigma}^{\dagger} c_{j s' \sigma} ,
\end{eqnarray}
and
\begin{eqnarray}
H_{I} = U \sum_{i s } n_{i s \uparrow} n_{i s \downarrow} .
\end{eqnarray}
Here, $c_{is\sigma}$ ($c_{is\sigma}^{\dagger}$) describes annihilation (creation) of an electron with spin $\sigma \in \{ \uparrow , \downarrow \}$ in the {\it i}-th site of sublattice $s \in \{ A , B \}$ (Fig.~\ref{fig.latt}.a). 
The first term describes a non-interacting state.
We assume equal hopping between the nearest-neighbor (NN) sites (i.e.\ $t_{ij}^{ss'} = t = 1$ as energy unit and 0 otherwise).
$\mu$ is the chemical potential, whereas $h$ the external magnetic field. 
The second term describes the on-site Coulomb interaction $U/t < 0$ being the source of {\it s-wave} type superconductivity.

\subsection{Non-interacting state}

In the absence of interaction ($U = 0$), the kinetic term~(\ref{eq.hkin}) can be transformed to the reciprocal space as follows:
\begin{eqnarray}
H_{K} &=& \sum_{{\bm k},s,\sigma} \left( - \mu - \sigma h \right) c_{{\bm k}s\sigma}^{\dagger} c_{{\bm k}s\sigma} \\
\nonumber &+& \sum_{{\bm k},\sigma} - t \left( g ( {\bm k} ) c_{{\bm k}A\sigma}^{\dagger} c_{{\bm k}B\sigma} + h.c. \right) ,
\end{eqnarray}
where $g ({\bm k}) = \sum_{i=1}^{3} \exp ( i {\bm k} \cdot {\bm \delta}_{i} )$. 
Here, ${\bm \delta}_{i}$ defines the location of the NN sites (Fig.~\ref{fig.latt}.a).
Hence, one obtains:
\begin{eqnarray}
\nonumber g ( {\bm k} ) &=& \sqrt{ 3 + 2 \cos \left( \sqrt{3} k_{x} \right) + 4 \cos \left( \frac{ \sqrt{3} }{2} k_{x} \right) \cos \left( \frac{3}{2} k_{y} \right) } . \\
\end{eqnarray}
Using the Nambu notation, $H_{K}$ can be rewritten in the following way:
\begin{eqnarray}
H_{K} = \sum_{{\bm k}\sigma} \Phi_{{\bm k}\sigma}^{\dagger} \mathbb{H} ({\bm k},\sigma) \Phi_{{\bm k}\sigma} ,
\end{eqnarray}
where $\Phi_{{\bm k}\sigma}^{\dagger} = \left( c_{{\bm k}A\sigma}^{\dagger} , c_{{\bm k}B\sigma}^{\dagger} \right)$ is the Nambu spinor, and
\begin{eqnarray}
\label{eq.hamkin.nosc} && \mathbb{H} ({\bm k},\sigma) = 
\left(
\begin{array}{cc}
- ( \mu + \sigma h ) & - t g ( {\bm k} ) \\ 
- t g ( {\bm k} ) & - ( \mu + \sigma h ) \\ 
\end{array}
\right) .
\end{eqnarray}
The eigenvalues $\mathcal{E}_{\alpha{\bm k}\sigma}$ of the Hamiltonian $H_{K}$ can be found by diagonalization of the matrix~(\ref{eq.hamkin.nosc}). 
As a result, one obtains:
$\mathcal{E}_{\pm,{\bm k}\sigma} = \pm t | g ({\bm k}) | - ( \mu + \sigma h )$ (Fig.~\ref{fig.latt}.b).

\subsection{Superconducting state}

The source of the {\it s-wave} superconductivity in the Hubbard model is the on-site attraction ($U/t < 0$) between particles with opposite spins on the same site. 
The interaction term $H_{I}$ can be decoupled in the mean field approximation by:
\begin{eqnarray}
\label{eq.mfa_op} n_{is\uparrow} n_{is\downarrow} = \Delta_{i,s}^{\ast} c_{is\downarrow} c_{is\uparrow} + \Delta_{i,s} c_{is\uparrow}^{\dagger} c_{is\downarrow}^{\dagger} - | \Delta_{i,s} |^{2} ,
\end{eqnarray}
where 
$\Delta_{i,s} = \langle c_{is\downarrow} c_{is\uparrow} \rangle$
is the superconducting order parameter (SOP) in the sublattice {\it s}.
Then,
\begin{eqnarray}
H_{I}^{MF} = U \sum_{is} \left( \Delta_{i,s} c_{is\uparrow}^{\dagger} c_{is\downarrow}^{\dagger} + H.c. \right) ,
\end{eqnarray}
where the last term from Eq.~(\ref{eq.mfa_op}) has been omitted, because it does not affect the self-consistent equations.
However, it is important to emphasize that this term has to be taken into account in a grand canonical potential calculation to determine the stability of different phases, since this constant term decreases the energy of the system~\cite{januszewski.ptok.15}.

Because there are two shifted sublattices (A and B) in the system, the SOP term for the FFLO phase can be rewritten as:
\begin{eqnarray}
\label{eq.sop} & \Delta_{j,s} & = \\
\nonumber && \Delta_{0} \left[ \delta_{s,A}
\exp ( i {\bm Q} \cdot {\bm R}_{j} ) + \delta_{s,B} \exp \left( i {\bm Q} \cdot ( {\bm R}_{j} + {\bm w} ) \right) \right] ,
\end{eqnarray}
where $\Delta_{0}$ is the SOP amplitude in the entire system, whereas ${\bm Q}$ is the total momentum of the Cooper pair. 
Here, ${\bm R}_{i}$ denotes the location of the {\it i}-th site in real space, while ${\bm w}$ describes the shift between both atoms in the unit cell and equals ${\bm \delta}_{2}$ (cf.\ Fig.~\ref{fig.latt}.a). 
In the superconducting phase ($\Delta_{0} > 0$), one can distinguish the BCS state with $| {\bm Q} | = 0$ and the FFLO phase for $| {\bm Q} | > 0$.
Hence, in momentum space:
\begin{eqnarray}
H_{I}^{MF} &=&  U \sum_{\bm k} \Delta_{0} \left( c_{{\bm k}A\uparrow}^{\dagger} c_{-{\bm k}+{\bm Q}A\downarrow}^{\dagger} \right. \\
\nonumber && + \left. \exp \left( i {\bm Q} \cdot {\bm w} \right) c_{{\bm k}B\uparrow}^{\dagger} c_{-{\bm k}+{\bm Q}B\downarrow}^{\dagger} \right) + H.c.
\end{eqnarray}
As a consequence, the mean field Hamiltonian $H^{MF} = H_{K} + H_{I}^{MF}$ 
can be rewritten in a block matrix form: 
\begin{eqnarray}
H^{MF} = \sum_{\bm k} \Psi_{\bm k}^{\dagger} \mathbb{H}_{MF} ({\bm k}) \Psi_{\bm k},
\end{eqnarray}
where $\Psi_{\bm k}^{\dagger} \equiv \left( \Phi_{{\bm k}\uparrow}^{\dagger} , \Phi_{-{\bm k}+{\bm Q}\downarrow}^{T} \right)$, while the partial block matrix $\mathbb{H}_{\bm k}^{MF}$ is given as:
\begin{eqnarray}
\mathbb{H}_{\bm k}^{MF} = \left(
\begin{array}{cccc}
\mathbb{H} ({\bm k},\uparrow) & \mathbb{U} ({\bm Q}) \\
\mathbb{U}^{\ast} ({\bm Q}) & -\mathbb{H}^{\ast} (-{\bm k}+{\bm Q},\downarrow) \\
\end{array} 
\right) .
\label{eq.ham.matrix.block}
\end{eqnarray}
The diagonal elements of $\mathbb{H}_{\bm k}^{MF}$, i.e.\ ones involving the matrix $\mathbb{H} ({\bm k},\sigma)$, describe the single-particle spectrum and are given by Eq.~(\ref{eq.hamkin.nosc}), while the off-diagonal elements describe superconductivity and $\mathbb{U}(\bm Q)$ is defined as $\mathbb{U}({\bm Q}) = U \Delta_{0} \delta_{ss'} \left( \delta_{s,A} + e^{ i {\bm Q} \cdot {\bm w} } \delta_{s,B} \right)$, where the index of matrix elements describes sublattices.

\section{Numerical results and discussion}
\label{sec.numres}

In this section we show and discuss the numerical results. 
First, we describe the details of numerical predictions (Sec.~\ref{sec.numres.details}). Next, we present the phase diagrams for the half-filling and non-half-filling
(i.e.\ doped) case (Sec.~\ref{sec.numresphasediagram}), and we discuss them in the context of the density of states analysis (Sec.~\ref{sec.numresdos}).
Finally, we provide the numerical calculations and discuss the main and novel properties of the FFLO phase in the hexagonal lattice (Sec.~\ref{sec.numrescpmom}).

\subsection{Numerical details}
\label{sec.numres.details}

To find the ground state, we calculate the grand canonical potential, defined by $\Omega \equiv - k_{B} T \ln \{ \Tr [ \exp ( - H^{MF} / k_{B} T ) ] \}$, which at $T=0$ is equivalent to the mean-field energy.
The calculations were performed in momentum space, on a $N = 121 \times 121$ ${\bm k}$-grid inside the first Brillouin zone (FBZ).
Since we study the stability of the FFLO phase, $\Omega$ is a function of the SOP amplitude $\Delta_{0}$ and the total momentum ${\bm Q}$ of Cooper pairs~\cite{ptok.cichy.17}. 
In this case, the procedure of minimization of $\Omega$ with respect to the SOP amplitude $\Delta_{0}$ and all possible momenta ${\bm Q}$ realized in the system is essential. To find the global minimum of $\Omega ( \Delta_{0} ,  {\bm Q} )$, the numerical calculations were accelerated using Graphics Processing Units (GPUs), according to the procedure described in Ref.~\cite{januszewski.ptok.15}.

It is important to emphasize that the mean-field approximation (MFA) overestimates, in general, the critical temperatures and the range of stability of the phases with a long-range order.
However, MFA gives at least qualitative description of the system
in the ground state ($T=0$), even in the strong coupling limit~\cite{micnas.ranninger.90}.

The ansatz which we have proposed to describe the SOP in real space, i.e. Eq.~(\ref{eq.sop}), does not limit the solutions with respect to the Cooper pairs momentum ${\bm Q}$.
It is a very important extension in comparison to the previous theoretical works in which the assumed ansatz strongly limits the possibility of the stable phase occurrence. 
For instance, it is worth to mention the {\it Kekul{\'e} order}~\cite{klein.hite.86}, for which the SOP in real space is $2\pi/3$ phase modulated~\cite{roy.herbut.10,li.jiang.17}.
Moreover, using the ansatz proposed in our paper, one can provide the analysis of phases other than FFLO, e.g. the spatially homogeneous spin-polarized superfluidity (called breached pair state or Sarma phase~\cite{sarma.63}). However, our numerical calculations show that this type of phase is unstable, for the whole region of parameters, which is in agreement with other theoretical works~\cite{hu.liu.06,liu.hu.07,devreese.klimin.11}.
Moreover, using this ansatz, e.g. pairing in the presence of the Fermi surface deformation~\cite{oganesyan.kivelson.01,barci.oxman.06,barci.bonfim.13,schlottmann.17} (called {\it Pomeranchuk instability}~\cite{pomeranchuk.59}) or multiparticle instability~\cite{whitehead.conduit.18} can be analyzed. 
However, these types of unconventional phases go beyond the scope of this work.

Additionally, the existence of a discontinuous phase transition between the BCS and the FFLO phase or the normal state, which is characteristic for the systems in the Pauli limit, leads to the occurrence of the phase separation regions.
In contrast to the case of a fixed chemical potential, if the number of particles is fixed, one obtains two critical Zeeman fields in the phase diagrams which determine the phase separation (PS) region between different types of phases~\cite{bedaque.caldas.03,sheehy.radzihovsky.06,sheehy.radzihovsky.07,cichy.micnas.11,cichy.micnas.14,ptok.cichy.17}, e.g. the BCS and the FFLO phase or the normal state.

\subsection{Phase diagram} 
\label{sec.numresphasediagram}

\begin{figure}[!t]
\centering
\includegraphics[width=0.75\linewidth]{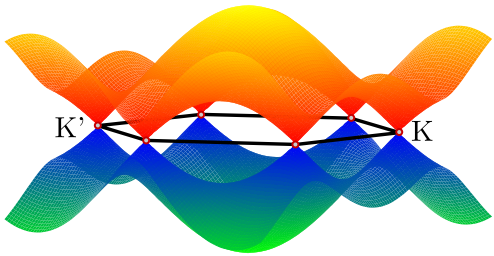}
\caption{
\label{fig.bands}
Energy band structure of the honeycomb lattice.
The first Brillouin zone (FBZ) is shown by the black hexagon. 
Simultaneously, the hexagon shows the Fermi level in the case of half filling ($\mu/t = 0$) and absence of the magnetic field ($h / t = 0$). 
The {\it Dirac cones} are located in the corner K and K' points of the FBZ.
}
\end{figure}

\begin{figure}[!b]
\centering
\includegraphics[width=\linewidth]{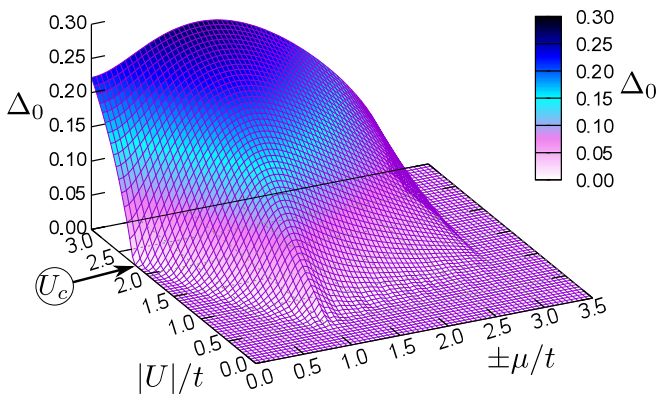}
\caption{
\label{fig.qpt}
Amplitude of the superconducting order parameter in the absence of the Zeeman magnetic field as a function of the chemical potential $\mu$ and pairing interaction $U$. 
}
\end{figure}

In the normal state, based on the dispersion relation $\mathcal{E}_{\alpha{\bm k}\sigma}$, one can distinguish the conduction ($\alpha = +$) and valence ($\alpha = -$) bands in the band structure of the system. 
At half-filling (for $\mu/t=0$, for which the average number of particles per lattice site $n = \frac{1}{N} \sum_{is\sigma} \langle c_{is\sigma}^{\dagger} c_{is\sigma} \rangle = \frac{1}{N} \sum_{{\bm k}s\sigma} \langle c_{{\bm k}s\sigma}^{\dagger} c_{{\bm k}s\sigma} \rangle$ is equal $1$) and at $h/t=0$, the conduction/valence band is fully empty/occupied, and the system exhibits a semi-metal behavior (Fig.~\ref{fig.bands}).
These two bands touch each other at the corner points of the first Brillouin zone (FBZ) in the {\it Dirac cones}, which is manifested by vanishing DOS at the Fermi level.

\begin{figure}[!t]
\centering
\includegraphics[width=\linewidth]{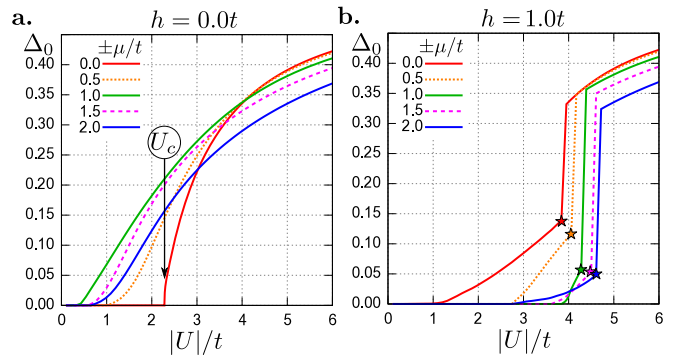}
\caption{
\label{fig.sop}
Amplitude of the superconducting order parameter $\Delta_{0}$ as a function of the chemical potential $\mu$ and attractive interaction $U$.
The results in the absence (a) and in the presence (b) of the magnetic field $h$.
Stars at panel b show the phase transition from FFLO to BCS with increasing $|U|$.
}
\end{figure}

At half-filling ($\mu / t = 0$, $n = 1$) and in the absence of the magnetic field, the honeycomb lattice exhibits a continuous quantum phase transition between semi-metal phase and the BCS state~\cite{zhao.paramekanti.06,seki.ohta.12,lin.liu.15} (Fig.~\ref{fig.qpt}).
The superconducting state can emerge in the system for pairing interaction stronger than some critical interaction $U_{c}$.
We estimate $|U_{c}|/t \simeq 2.245$ (Fig.~\ref{fig.sop}.a), which is in good agreement with previous mean-field studies~\cite{zhao.paramekanti.06}.
However, for any $\mu \neq 0$, the SOP exhibits exponential-like decrease to zero with decreasing $| U |$ (Fig.~\ref{fig.sop}.a).
This behavior is well visible around VHS ($\mu/t=\pm 1$ at Fig.~\ref{fig.qpt}).

The increase of the attraction above $U_{c}$ leads to the stabilization of the BCS state (Fig.~\ref{fig.sop}.a).
With an increasing Zeeman magnetic field (increasing population imbalance), the FFLO phase becomes stable at some $|U|$-dependent critical value $h_{c}$, through a first order phase transition.
The discontinuous phase transition is manifested by a jump of the order parameter with increasing $|U|$ and at fixed $h$, which is illustrated in Fig.~\ref{fig.sop}.b and indicated by stars. 
As we mentioned above, this behavior of the order parameter is reflected in the occurrence of the phase separation region in the phase diagrams for fixed $n$.
Indeed, such behaviour is observed in the system under consideration as well, which will be disucssed in detail in the next paragraph.

The essential finding of our work is that the FFLO phase can also be stable below $U_{c}$, for some range of magnetic fields.
As we already emphasized, this feature makes the experimental realization of this phase much simpler because any superconducting state which appears in the range $0 > U > U_{c}$ can only be the FFLO phase.
Preparing the experimental setup in such a way that the average number of particles per lattice site is equal to one, while introducing a mismatch between the atoms with ``\emph{up}'' and ``\emph{down}'' spins, and tuning the interaction to be between $U=0$ and $U_{c}$, facilitates observing and identifying FFLO phase (see also some remarks in the last paragraph of this section).

\begin{figure}[!t]
\centering
\includegraphics[width=0.75\linewidth]{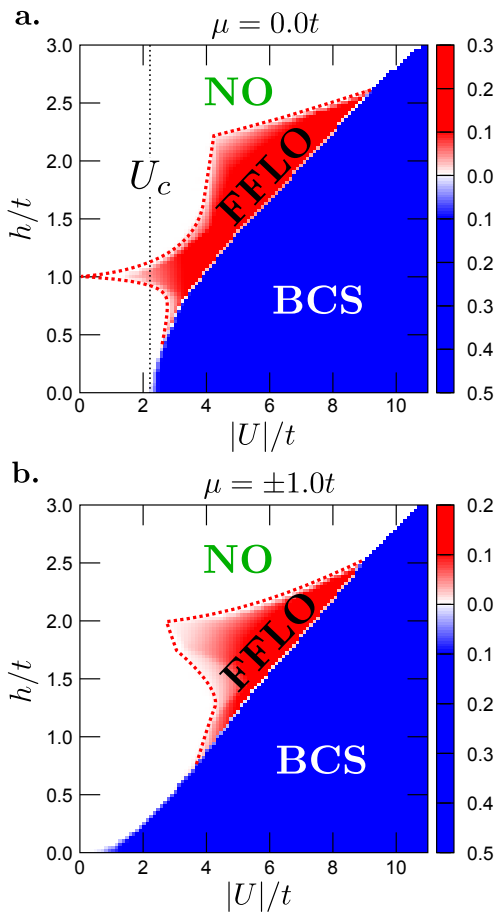}
\caption{
\label{fig.df}
Ground state phase diagram; the magnetic field $h$ vs.\ the attractive interaction $U$ at $\mu/t = 0$ (a) and $\mu/t = \pm 1$ (b).
The color map shows the SOP amplitude $\Delta_{0}$ (blue/red color for BCS and FFLO phases, respectively), whereas white color indicates the normal (NO) phase. At $\mu/t=0$ ($n=1$), the reentrant FFLO superconductivity is stable around $h/t=1$, below $U_c$.
}
\end{figure}

\begin{figure}[!t]
\centering
\includegraphics[width=\linewidth]{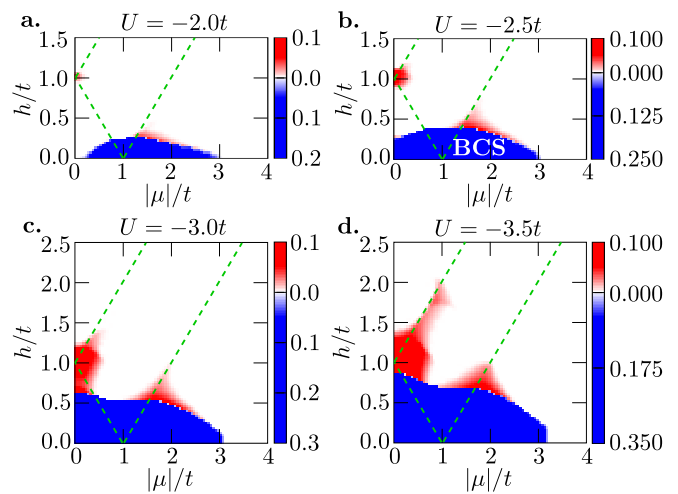}
\caption{
\label{fig.df2}
Ground state phase diagram; magnetic field $h$ vs.\ chemical potential $\mu$ for different values of the interaction $U$.
The color map shows the SOP amplitude $\Delta_{0}$ (blue/red color for BCS/FFLO phases, respectively, white color indicates the NO phase). 
Green dashed lines indicate parameters for which the VHS are located at the Fermi level.
}
\end{figure}

\begin{figure}[!b]
\centering
\includegraphics[width=\linewidth]{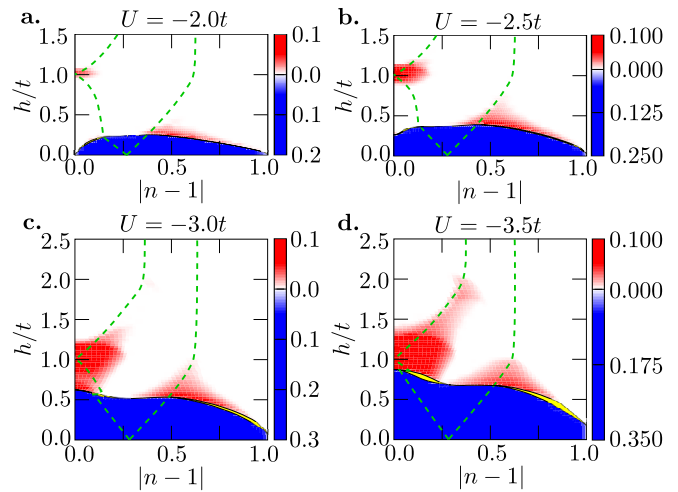}
\caption{
\label{fig.df3}
Ground state phase diagram; magnetic field $h$ vs.\ avarage number of particles $n$. 
The color map shows the SOP amplitude $\Delta_{0}$ (blue/red color for BCS/FFLO phases, respectively, white color indicates the NO phase.
Yellow area denotes the phase separation region. Green dashed lines indicate parameters for which the VHS are located at the Fermi level.
}
\end{figure}

If the chemical potential and hence the density is changed, the character of the phase diagram changes (cf. Fig.~\ref{fig.qpt} and Fig.~\ref{fig.df} for $h/t = 0$).
As mentioned above, in the absence of a Zeeman magnetic field and at half-filling, the system exhibits a quantum phase transition. 
As a consequence of this fact, superconductivity becomes unstable above some critical value of attraction ($|U_{c}|$) (shown in Fig.~\ref{fig.qpt}, Fig.~\ref{fig.sop}.a and Fig.~\ref{fig.df}.a).
However, at any small deviation from half-filling (i.e. for any non-zero doping), the superconductivity is stable for the whole range of attractive interactions and one can observe an exponential decay of the order parameter with decreasing $|U|$ (e.g. Fig.~\ref{fig.qpt} or Fig.~\ref{fig.df}.b shows the case of $\mu/t = \pm 1$).
Away from half-filling and for small values of the attraction ($|U|/t\lesssim 4$), the FFLO phase occurs for larger values of $h$ than in the half-filled system.
It is important to emphasize that the phase transition from the BCS to the normal state as well as from the BCS to the FFLO phase is always discontinuous. However, the phase transition from the FFLO to the normal state is of second order, for the whole range of parameters.
Hence, the FFLO phase, in comparison to the BCS state, evinces a {\it reentrant} behavior (i.e. appear and then disappear when varying $h$ at fixed $U$ and $t$), because the FFLO phase can occur at $h > 0$ for some $|U| < | U_{c} |$, even without the BCS as a {\it precursor} at half-filling.
These properties are novel and have not been described in the literature so far.

In both cases, i.e.\ at half-filling and away from it, the boundaries of the FFLO and BCS phases (critical magnetic fields) show typical behavior at larger values of $U$~\cite{ptok.cichy.17}, i.e.\ the FFLO state becomes unstable with an increasing attractive interaction because of the vanishing of at least one Fermi surface~\cite{conduit.conlon.08,heidrichmeisner.feiguin.10}. In this case, the system is in a phase of tightly bound local pairs (hard-core bosons)~\cite{micnas.ranninger.90,cichy.micnas.11,cichy.micnas.14}.

The influence of the presence of VHS in DOS on the stabilization of the FFLO phase is illustrated in Fig.~\ref{fig.df2} with $h-\mu$ phase diagrams, at four fixed values of $U$. 
Green dashed lines indicate parameters for which the VHS are located at the Fermi level. Around these lines, for larger values of magnetic fields, the FFLO state becomes stable. 
Moreover, the evolution of the reentrant transition with $U$, at $\mu=0$, is clearly visible.
It should also be emphasized that similar results to those presented in this paper, for Lieb and Kagome lattices, can be reproduced using more advanced methods like e.g. the Dynamical Mean-Field Theory (DMFT)~\cite{huhtinen.tylutki.18}.

As mentioned above, discontinuous phase transition can lead to the occurrence of a phase separation in the case of a fixed number of particles $n$. 
It can be found by mapping of phase diagrams at a fixed chemical potential $\mu$ (Fig.~\ref{fig.df2}) onto the phase diagrams with fixed $n$ (Fig.~\ref{fig.df3}).
The region of parameters for which the phase separation is observed is shown in Fig.~\ref{fig.df3} (the yellow area).
The existence of the FFLO phase leads to the suppression of the phase separation region. Hence, it is more likely to observe the PS region rather between spatially homogeneous phases such as the BCS and the normal states.

\begin{figure}[!b]
\centering
\includegraphics[width=\linewidth]{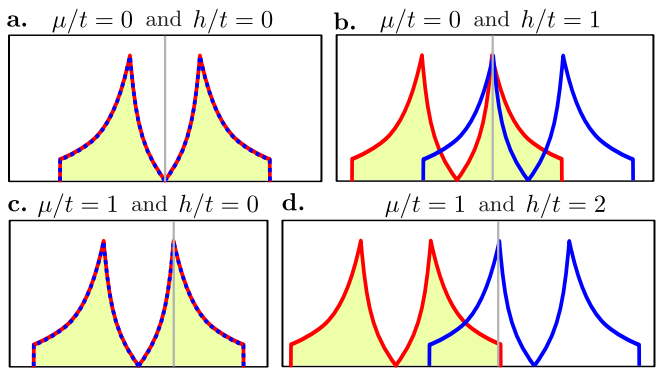}
\caption{
\label{fig.dos}
DOS of the honeycomb lattice ($U=0$) for different values of the chemical potential $\mu$ and external magnetic field $h$. 
Red and blue lines denote DOS for particles with spin $\uparrow$ and $\downarrow$, respectively, whereas the Fermi level is shown as gray line. 
The scheme for different parameters $(\mu/t, h/t)$: (0,0), (0,1), (1,0) and (1,2) (panels a-d).
}
\end{figure}

One needs to remember that FFLO phases are known to be much more sensitive to thermal fluctuations than the BCS state, and typically have very low critical temperatures.
Hence, the experimental detection of these phases could be still rather problematic.
Moreover, for a two-dimensional system, at a zero Zeeman magnetic field, the superconducting-normal transition in the attractive Hubbard model is of the Kosterlitz-Thouless (KT) type, mediated by unbinding of vortices, i.e. below the KT temperature, the system has a quasi-long-range (algebraic) order, which is characterized by a power law decay of the order parameter correlation function and a non-zero superfluid stiffness.
As has been shown in Ref.~\cite{cichy.micnas.11}, the KT phase (quasi-superconducting Sarma phase for a homogeneous system) is restricted to the weak coupling region and low values of polarizations (magnetic fields).

\subsection{Density of states analysis}
\label{sec.numresdos}

DOS of the honeycomb lattice shows $1/\sqrt{E}$ singularities due to the one-dimensional nature of the electronic spectrum~\cite{castroneto.guinea.09,saito.fujita.92,saito.fujita.92b,gonzalez.08}. 
Moreover, near the ``neutral'' point ($E=0$), DOS can be approximated by $\rho ( \omega ) \propto | \omega |$. 
As we show below, the presence of the VHS in DOS, located at $\omega/t = \pm 1$, at half-filling ($\mu/t = 0$) and $h/t = 0$ (e.g.\ Fig.~\ref{fig.dos}.a), is important from the point of view of unconventional superconductivity. 
As a consequence of the existence of two equivalent sublattices, there are two VHS in DOS.  
Changing the location of the Fermi level by changing the value of the chemical potential $\mu$ (filling $n$) or external magnetic field $h$ in the system, one can change the relative position between VHS for particles with spin ``\emph{up}'' ($\uparrow$) and ``\emph{down}'' ($\downarrow$) (red and blue lines in Fig.~\ref{fig.dos}, respectively).

\begin{figure}[!b]
\centering
\includegraphics[width=0.6\linewidth]{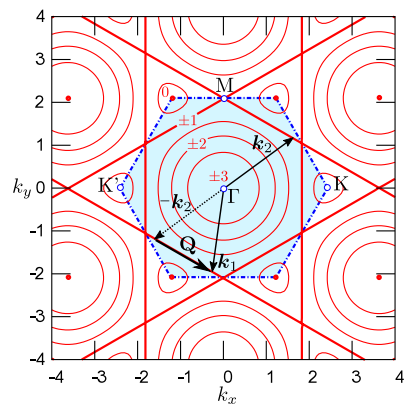}
\caption{
\label{fig.fs}
The Fermi surface of the honeycomb lattice for different values of the Fermi level is shown by isoenergetic red lines.
The first Brillouin zone is shown by a blue hexagon.
Vectors show an example of the Cooper pair formation, while solid red line shows the Fermi level for the filling equal 3/8 and 5/8.
}
\end{figure}

It is important to emphasize that the DOS has influence on the critical temperature. In the BCS theory: $T_{c} \propto \exp \left( - 1 / | U | \rho ( E_{F} ) \right)$, where $ \rho ( E_{F} )$ is the total DOS at the Fermi level $E_{F}$ for both spin components.
We describe the behavior of DOS schematically with the example shown in Fig.~\ref{fig.dos}, in relation to some characteristic parameters taken from the phase diagram in Fig.~\ref{fig.df}.
Without magnetic field ($h=0$), at half-filling ($\mu = 0$), $E_{F}$ is located at the neutral points $K$ and $K'$, with $E = 0$ (Fig.~\ref{fig.dos}.a).
Consequently, there exists a critical value of the interaction $U_{c}$ below which the BCS phase is unstable and the normal state is favored (the semi-metal-superconductor transition) (see Fig.~\ref{fig.df}.a). 
A similar phenomenon is also observed e.g.\ in the metal-insulator transition~\cite{herbut.juricic.09}.
In the presence of a Zeeman magnetic field, the DOSs are unequal for the particles with opposite spins.
For instance, at $h/t = 1$, DOSs 
are shifted in the way illustrated in Fig.~\ref{fig.dos}.b. 
Both VHSs are located at $E_{F}$, with energies $\omega/t = \pm 1$, where $+$/$-$ corresponds to particles with spin $\uparrow$/$\downarrow$, respectively. 
Consequently, DOS has a maximum at $E_{F}$. 
The large spin-imbalance implicates the stabilization of the FFLO phase.
Similar behavior can be observed in the case of an over-/underdoped system  (e.g.\ $\mu/t =\pm 1$) without magnetic field (Fig.~\ref{fig.dos}.c). 
In this case, both VHS (for both spin components) with energies $\omega/t = \pm 1$ are located at $E_{F}$, whereas spin imbalance does not exist.
Consequently, the BCS phase is stable.
Hence, the superfluid phase can be realized for any pairing interaction strength because of the finite value of $\rho ( E_{F} )$. 
If the magnetic field is increased, DOS is shifted again. For $h/t = 2$, only VHS for particles with spin up is located at $E_{F}$ (Fig.~\ref{fig.dos}.d). 
In this case, i.e.\ for large magnetic fields, the attractive interaction can lead to the stabilization of the FFLO phase.
However, it is important to emphasize that there is a critical value of $U$ below which the FFLO state becomes unstable, in contrast to the half-filled case.


\begin{figure}[!b]
\centering
\includegraphics[width=\linewidth]{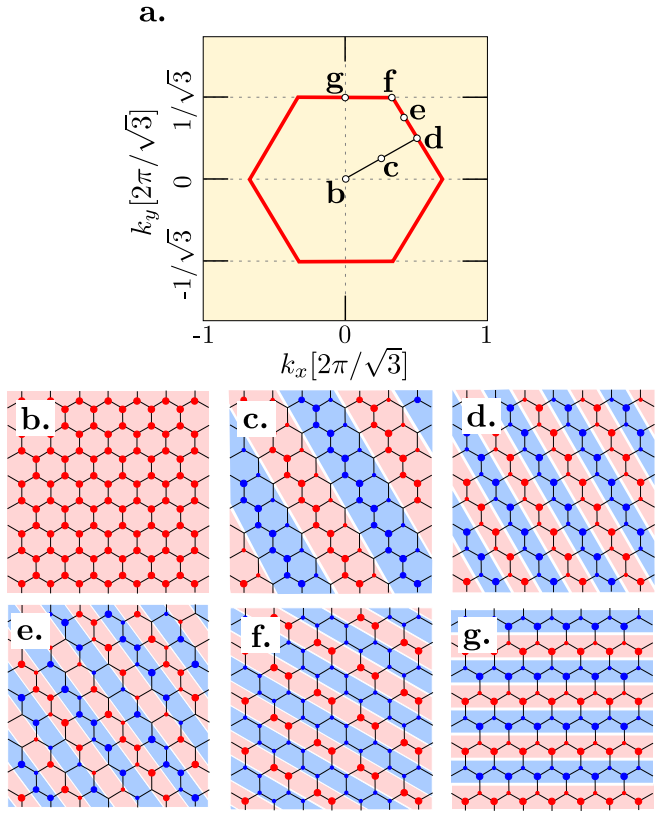}
\caption{
\label{fig.mod}
The spatial decomposition of the SOP in real space for different total momenta of Cooper pairs ${\bm Q}$ (marked by points b-g in panel a).
Color (red/blue) and the size of the circles correspond to the sign ($+$/$-$) and the value of the order parameter.
White lines are nodal lines in real space.
}
\end{figure}

As mentioned above, the mutual position of the DOSs for particles with opposite spins is crucial for the stabilization of the BCS state as well as the FFLO phase. 
For instance, to stabilize the FFLO state, the system should be doped to the so-called $M$ point of FBZ.      
This situation corresponds to a 3/8 or 5/8 filling in a given spin-type band~\cite{nandkishore.levitov.12,nandkishore.thomale.14}.
At these fillings, VHS originates from three non-equivalent saddle points.
Moreover, the Fermi surface exhibits a high degree of nesting (Fig.~\ref{fig.fs}), forming a perfect hexagon at this filling~\cite{gonzalez.08}).
These two features lead to the stabilization of the FFLO phase, as a consequence of the perfect nesting of the Fermi surfaces corresponding to the opposite spins~\cite{shimahara.94,ptok.17}.
It can be described using notation from Fig~\ref{fig.fs}.
In the case of the mentioned filling (i.e. 3/8 and 5/8 filling), the Fermi surfaces for the particles with spin up and down are degenerate (shown by solid red line). 
Hereby, the Cooper pairs with total momentum ${\bm Q}$ can be formed by the particles with momentum ${\bm k}_{1}$ and ${\bm k}_{2}$.
Because of the fact that the Fermi surface is given by the hexagon, the Cooper pairs with momentum ${\bm Q}$ (along $\Gamma-M$ line) can be realized for many different ${\bm k}_{1}$ and ${\bm k}_{2}$.
Then, the FFLO state can be energetically more favorable for larger range of the pairing interaction $U$.
The situation described above is clearly visible e.g. in Fig.~\ref{fig.df}.a, at $h/t = 1$.

\begin{figure}[!b]
\centering
\includegraphics[width=\linewidth]{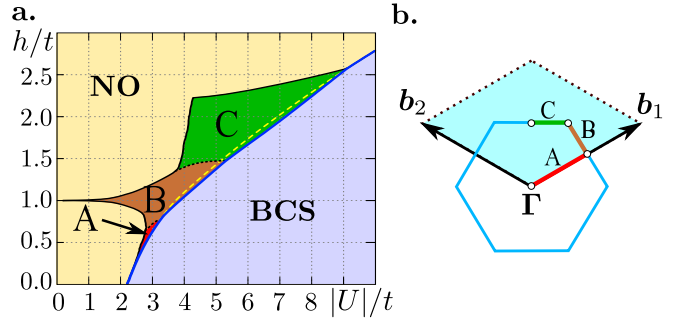}
\caption{
\label{fig.tmcp}
(a) The mean field ground state phase diagram. The magnetic field $h$ vs. pairing interaction $U$ at half-filling ($\mu/t = 0$).
In the case when the FFLO phase is neglected in calculations, the dashed red line indicates the critical magnetic field above which the BCS state is unstable.
Above this line, the normal state (NO) exists. 
The occurrence of the FFLO state in the phase diagram slightly shifts the boundary of the BCS phase which is indicated by the solid blue line. 
The labels A, B, C show three different directions of the total momentum ${\bm Q}$ for which the FFLO phase is realized.
(b) Schematic picture of vectors ${\bm Q}$ in FBZ, for three different variants of the FFLO phase: A, B and C.
}
\end{figure}

\begin{figure}[!t]
\centering
\includegraphics[width=0.75\linewidth]{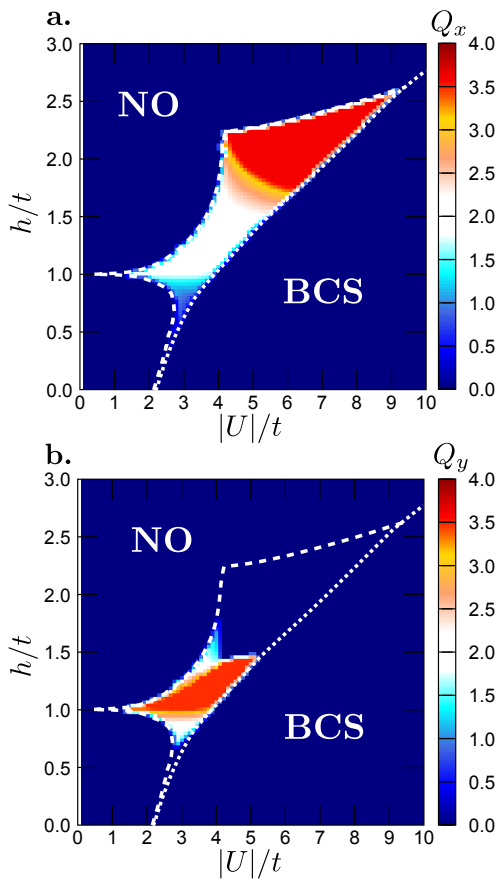}
\caption{
\label{fig.cp_mom}
Components of the Cooper pairs momentum vector ${\bm Q} = ( Q_{x} , Q_{y} )$ as a function of the interaction strength $U$ and the external magnetic field $h$.
}
\end{figure}

\subsection{Cooper pairs momentum $\bm Q$ dependence}
\label{sec.numrescpmom}

The dependence of the Cooper pairs properties is also related to the nesting of Fermi surfaces. 
This is clearly visible in the evolution of the total momentum ${\bm Q}$ of pairs with increasing magnetic field (Fig.~\ref{fig.mod}).
Usually (for instance, in the square~\cite{januszewski.ptok.15,ptok.15} or triangular~\cite{guo.jiang.11,ye.pan.10,ptok.maska.09} lattice case), only the length of ${\bm Q}$ changes, without changing the direction.
It is the consequence of the mutual shift of the Fermi surfaces for the particles with spin up and down. 
Moreover, the direction of ${\bm Q}$ can be found, within good approximation, from the Cooper pairs susceptibility calculation~\cite{ptok.crivelli.13,januszewski.ptok.15,ptok.kapcia.17}.
However, it is worth to emphasize that only the global minimum of the energy with respect to ${\bm Q}$ and $\Delta_{0}$ has to be found to give proper information on the BCS/FFLO phase.

In the case of the honeycomb lattice, with two atoms per unit cell, ${\bm Q}$ is not subject to the typical evolution described above.
Instead, the evolution of ${\bm Q}$ with increasing magnetic field can be divided into three phases (shown in Fig.~\ref{fig.tmcp}):
\\
{\it (A)} the evolution along the reciprocal lattice vectors,
\\
{\it (B)} the evolution along the boundary of the FBZ, perpendicular to reciprocal lattice vectors
and
\\
{\it (C)} the evolution along the boundary of the FBZ perpendicular to ${\bm w}$, which describes the mutual shift of two sublattices.
This evolution is a consequence of the nesting between the Fermi surfaces for particles with spin up and down, which are shifted by the Zeeman magnetic field.
As a consequence, the magnitude as well as the direction of ${\bm Q}$ change in a non-monotonic way with an increasing magnetic field $h$~(Fig.~\ref{fig.cp_mom}).
One can indicate the boundaries in the phase diagram between the FFLO phase with different directions of ${\bm Q}$.
The properties described above are shown in Fig.~\ref{fig.mod} and in Supplemental Material~\cite{mov.mom}, which schematically present the spatial decomposition of the SOP for different ${\bm Q}$ [in SM -- the small black crosses denote the position of lattice sites in real space, while the size and color of solid circles correspond to the value and sign of the SOP (blue and red color denote signs minus and plus, respectively)].

\section{Summary}
\label{sec.summary}

The honeycomb lattice exhibits a characteristic band structure in which two bands touch each other at the {\it Dirac cones} vertices.
Consequently, at half-filling, there exists some critical interaction $U_{c}$ above which the BCS phase becomes stable. 
This value indicates the occurrence of a quantum phase transition from the semi-metallic to the superconducting phase, in the absence of a Zeeman magnetic filed.
In this paper, we demonstrated that the behavior of the system changes significantly when population imbalance is introduced.
Such a system can be realized in ultracold gases experiments, by loading atoms in two different hyperfine states onto a honeycomb-shaped lattice.
In such case, the FFLO state with non-zero total momentum of Cooper pairs can be realized.
The characteristic features of the honeycomb lattice DOS can lead to the FFLO phase stabilization for any pairing interaction strength.
Moreover, at half-filling, $n = 1$, the FFLO phase shows a reentrant behavior.
For any pairing interaction (also lower than $U_{c}$), this phase can be realized without the BCS phase as a {\it precursor}, which is not observed in case of other lattices, e.g.\ square or triangular.
We explain this behavior as a consequence of the singular DOS and the strong nesting of Fermi surfaces.
These results can be helpful for experimental realization of the FFLO phase on an artificial hexagonal optical lattice, because any superconducting state which appears in the range of $0 > U > U_{c}$ can only be the FFLO phase.
Additionally, we show that the evolution of the total momentum of Cooper pairs is untypical. 
As a consequence of the nesting between the Fermi surfaces  for particles with different spins, the momenta change values and directions.

\begin{acknowledgments}
We thank Krzysztof Cichy, Peter G.\ J.\ van Dongen and Matteo Rizzi, for careful reading of the manuscript, valuable comments and discussions. We also thank Ravindra W. Chhajlany, Roman Micnas and Andrii Sotnikov for many fruitful discussions.
This work was supported by the National Science Centre (NCN, Poland) under grants: UMO-2017/24/C/ST3/00357 (A.C.) and UMO-2016/20/S/ST3/00274 (A.P.).
\end{acknowledgments}

\bibliography{biblio}

\end{document}